\RequirePackage{amsmath}

\documentclass{aip-cp}

\usepackage[numbers]{natbib}

\usepackage{graphicx,amsmath}
\usepackage{mathtools}

\begin{document}

% Title portion
\title{Solitary Wave Solution of Flat Surface
Internal Geophysical Waves with Vorticity}

\author[aff1,aff2]{Alan Compelli\corref{cor1}}
\eaddress[url]{http://www.maths.dit.ie}

\affil[aff1]{School of Mathematical Sciences, Dublin Institute of Technology, Kevin Street, Dublin 8, Ireland.}

\affil[aff2]{Erwin Schr\"odinger Institute for Mathematics and Physics,
University of Vienna, Vienna, Austria.}

\corresp[cor1]{Corresponding author: alan.compelli@dit.ie}

\maketitle

\begin{abstract}
A fluid system bounded by a flat bottom and a flat surface with an internal wave and depth-dependent current is considered. The Hamiltonian of the system is presented and the dynamics of the system are discussed. A long-wave regime is then considered and extended to produce a KdV approximation. Finally, a solitary wave solution is obtained. 
\end{abstract}

% Head 1
\section{INTRODUCTION}
Surface waves typically have heights of 1-2m with extreme wave heights greater than 20m having been observed, for example a 20.4m surface wave was measured off the northwest coast of Ireland in December 2011 by
Met \'{E}ireann (the Irish meteorological service). Internal waves are disturbances which act as an interface between discrete fluid bodies which have distinct properties such as salinity or temperature. As internal waves are typically 10 times higher than surface waves this heuristically suggests that extreme internal waves as high as 200m may occur. Indeed Alford \emph{et al.} \cite{bib1} have measured wave heights in excess of 170m in the Luzon Strait in the South China Sea. 

Studies of such gigantic waves are essential in the context of tsunami prediction, marine biology, the design of marine vessels, etc. 

Some studies in an irrotational \cite{bib2} and a rotational \cite{bib3} setting have been completed but the most pertinent is the long-wave and KdV approximations obtained in \cite{bib4}. The presented paper extends this analysis to produce a solitary wave solution.

\section{SETUP}
\begin{figure}[h]
  \centerline{\includegraphics[width=295pt]{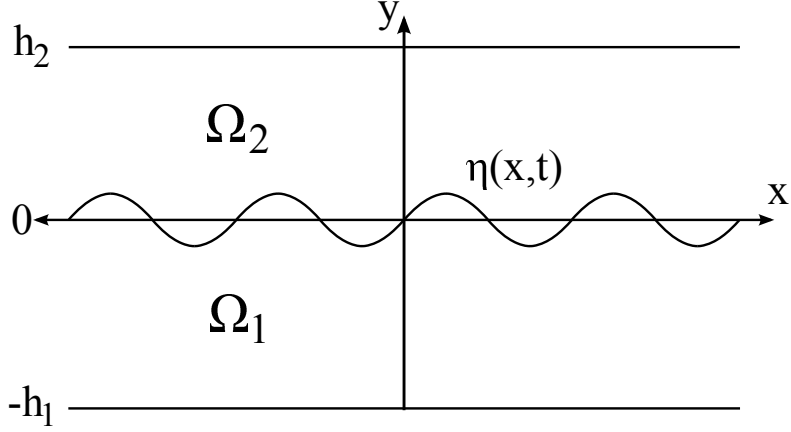}}
  \label{fig:Fig1}
  \caption{System setup.}
\end{figure}

A two-dimensional water wave system consisting of two discrete fluid domains separated by a free common interface in the form of an internal wave, such as a pycnocline or thermocline, is presented as per Figure \ref{fig:Fig1}. 

% Figure

The system is bounded at the bottom by an impermeable flatbed at a depth of $h_1$ and is considered as being bounded on the surface by an assumption of absence of surface motion by a lid at a height of $h_2$. The domains $\Omega_1=\{(x, y)\in\mathbb{R}^2: -h_1< y < \eta(x,t)\}$ and $\Omega_2=\{(x, y)\in\mathbb{R}^2: \eta(x,t)< y < h_2\}$ are defined with values associated with each domain using corresponding respective subscript notation 1 and 2. Propagation of the internal wave is assumed to be in the positive $x$-direction which is considered to be `eastward'. The centre of gravity is in the negative $y$-direction. 

The function $\eta(x,t)$ describes the elevation of the internal wave with the mean of $\eta$ assumed to be zero, $$\int\limits_{\mathbb{R}} \eta(x,t) dx=0.$$ 

The system is considered to be incompressible with $\rho_1$ and $\rho_2$ being the respective constant densities of the lower and upper media and stability is given by the immiscibility condition
\begin{equation}
\label{stability}
\rho_1>\rho_2.
\end{equation}

\begin{figure}[h]
  \centerline{\includegraphics[width=270pt]{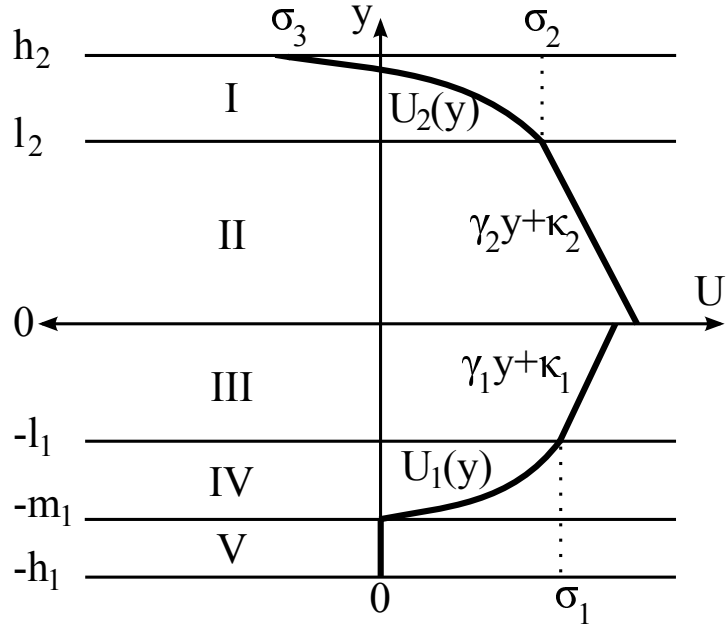}}
   \label{fig:Fig2}
  \caption{Current profile.}
\end{figure}

To avoid duplication of similar expressions in the 2 domains the subscript $i=\{1,2\}$ is introduced and so the velocity vector ${\bf{u}}_i=(u_i,v_i,0)$ is given in terms of stream functions $\psi_1,\psi_2$ as
\begin{equation}
    u_i=\psi_{i,y},\quad
    v_i=-\psi_{i,x}
\end{equation}
or in terms of wave-only velocity potentials $\widetilde{\varphi}_1,\widetilde{\varphi}_2$ and current $U(y)$ as
\begin{equation}
    u_i=  \widetilde{\varphi}_{i,x}+U(y),\quad
    v_i = \widetilde{\varphi}_{i,y}
\end{equation}
where the current profile is depicted by the heavy line in Figure \ref{fig:Fig2} consisting of 5 layers: layer I having an arbitrary current $U_2(y)$, layers II and III (which we will collectively refer to as the \emph{strip}, that is the layers adjacent to the wave) having linear profiles characterised by constant vorticities $\gamma_i$ and constant currents $\kappa_i$, layer IV having an arbitrary current $U_1(y)$ and layer V having zero current.

% Figure

It is noted at $y=0$ that the current is $\kappa_1$ from the perspective of $\Omega_1$ and $\kappa_2$ from the perspective of $\Omega_2$ and hence for $\kappa_1\neq\kappa_1$ there exists a vortex sheet.

The positive constants $\sigma_1-\sigma_3$ are included to demonstrate that a westward surface wind may be responsible for a westward drift experienced by observers on the surface but that this drift might reduce in strength and ultimately become an eastward drift for, say, a scuba diver descending in the presence of the considered current. This phenomenon was until relatively recently referred to (mysteriously) as \emph{dead water} and demonstrates how recent knowledge of such processes is.

We make the assumption that the functions $\eta(x, t),$ $\tilde{\varphi}_1(x, y, t)$ and $\tilde{\varphi}_2(x, y, t)$ belong to the Schwartz class $\mathcal{S}(\mathbb{R})$ with respect to the $x$ variable (for any $y$ and $t$). This reflects the localised nature of the wave disturbances. The assumption of course implies that for large absolute values of $x$ the internal wave attenuates and so
\begin{equation}
\lim_{|x|\rightarrow \infty}\eta(x,t)=0, \quad \lim_{|x|\rightarrow \infty}{ \tilde{\varphi}_1}(x,y,t)=0, \quad\lim_{|x|\rightarrow \infty}{ \tilde{\varphi}_2}(x,y,t)=0.
\end{equation}

The situation with a free surface is studied in \cite{bib5,bib6}.

\section{GOVERNING EQUATIONS}

The Bernoulli condition at the interface is \cite{bib4,bib5}
\begin{equation}
\rho_1\Big(({\varphi_{1,t}})_c+\frac{1}{2}|\nabla \psi_1|_c^2-(\gamma_1-2\omega)\chi_1 +g\eta\Big)=\rho_2\Big(({\varphi_{2,t}})_c+\frac{1}{2}|\nabla \psi_2|_c^2-(\gamma_2-2\omega)\chi_2+g\eta\Big)
\end{equation}
where
$\chi_i(x,t):=\psi_i\big(x,\eta(x,t),t\big)$ is the stream function evaluated at the interface, the subscript
$c$ means evaluation at the common interface (i.e. on the wave), 
$g$ is the acceleration due to gravity and $\omega$ is the rotational speed of Earth.

There is a kinematic boundary condition at the interface
\begin{equation}
 v_i=\eta_t +u_i\eta_x
\end{equation}
and boundary conditions at the bottom and top
\begin{equation}
    v_1(x,-h_1,t)=0,\quad
    v_2(x,h_2,t)=0
\end{equation}
respectively.

\section{HAMILTONIAN FORMULATION}
The Hamiltonian is given in terms of decomposed (wave and current) variables by the functional $H$ via
\begin{multline}
H\Big(\eta(x),\widetilde{\varphi}_i(x,y)\Big)= \frac{1}{2}\rho_1\int\limits_{\mathbb{R}} \int\limits_{-h_1}^{\eta(x)}|\nabla {\widetilde{\varphi}}_1|^2 dy dx+ \frac{1}{2}\rho_2\int\limits_{\mathbb{R}} \int\limits_{\eta}^{h_2} |\nabla {\widetilde{\varphi}}_2|^2 dy dx
+\rho_1\int\limits_{\mathbb{R}}  \int\limits_{-h_1}^{\eta(x)}  U\widetilde{\varphi}_{1,x} dy dx\\ +\rho_2\int\limits_{\mathbb{R}} \int\limits_{\eta}^{h_2} U \widetilde{\varphi}_{2,x} dy dx
+\frac{1}{2}\rho_1\int\limits_{\mathbb{R}}  \int\limits_{-h_1}^{\eta(x)} U^2  dy dx+\frac{1}{2}\rho_2\int\limits_{\mathbb{R}}  \int\limits_{\eta}^{h_2} U^2  dy dx+\frac{1}{2}(\rho_1-\rho_2)g\int\limits_{\mathbb{R}}\eta^2 dx.
\end{multline}

By introducing the interface velocity potentials
\begin{equation}
\xi_i(x):=\varphi_i(x,\eta(x)),
\end{equation}
the \emph{overall} interface velocity potential
\begin{equation}
\xi(x):=\rho_1\xi_1-\rho_2\xi_2,
\end{equation}
Dirichlet-Neumann operators (where $n_i$ are outward normals)
\begin{equation}
G_i(\eta)\xi_i:=\frac{\partial\varphi_i}{\partial n_i}\bigg |_{y=\eta}\sqrt{1+\big(\eta_x\big)^2},
\end{equation}
the operator $B$
\begin{equation}
B:=\rho_1 G_2(\eta)+\rho_2 G_1(\eta)
\end{equation}
and the variable $\mu$
\begin{equation}
\mu(x):=\big((\gamma_1-\gamma_2)\eta+(\kappa_1-\kappa_2)\big)\eta_x 
\end{equation}
the Hamiltonian can be written in terms of \emph{conjugate} variables as
\begin{multline}
\label{HamCONJ}
H\big(\eta(x),\xi(x)\big)=\frac{1}{2}\int\limits_{\mathbb{R}} \xi G_1(\eta) B^{-1}G_2(\eta)\xi \,dx
- \frac{1}{2}\rho_1\rho_2\int\limits_{\mathbb{R}}  \mu   B^{-1}\mu  \,dx
-\int\limits_{\mathbb{R}} (\gamma_1\eta+\kappa_1)\xi\eta_x \,dx\\+\rho_2\int\limits_{\mathbb{R}} \mu B^{-1}G_1(\eta)\xi\,dx
+\frac{\rho_1}{6\gamma_1}\int\limits_{\mathbb{R}}(\gamma_1 \eta+\kappa_1)^3dx-\frac{\rho_2}{6\gamma_2}\int\limits_{\mathbb{R}}(\gamma_2 \eta+\kappa_2)^3dx +\frac{1}{2}g(\rho_1-\rho_2 )\int\limits_{\mathbb{R}} \eta^2 dx
\end{multline}
and it is noted that (\ref{HamCONJ}) now describes the system in terms of wave quantities only.

The system has non-canonical equations of motion, analogous to those in \cite{bib7}:
\begin{equation}
        \eta_t=\frac{\delta H}{\delta \xi},\quad
        \xi_t=-\frac{\delta H}{\delta \eta} +\Gamma\chi 
\end{equation}
where, by establishing that $\chi_1=\chi_2$,
\begin{equation}
\chi:=\chi_1=\chi_2 
\end{equation}
and the constant $\Gamma$ is introduced as
\begin{equation}
\Gamma:=\rho_1\gamma_1-\rho_2\gamma_2+2\omega(\rho_1-\rho_2).
\end{equation}

It is noted, by the absence of $U_1$ and $U_2$ terms, that the dynamics of the system depends on the strip only.

We can use the variable transformation \cite{bib8}
\begin{equation}
\xi\rightarrow\zeta=\xi-\frac{\Gamma}{2} \int_{-\infty}^{x} \eta(x',t)\,dx',
\end{equation}
as $\xi$ is defined modulo an additive constant, and use the fact that
\begin{equation}
\int\limits_{\mathbb{R}}\eta(x,t) \,dx=\mbox{constant}
\end{equation}
(noting that the variable $\zeta(x,t)$ also belongs to the Schwartz class $\mathcal{S}(\mathbb{R})$ with respect to $x$) to give canonical equations of motion
\begin{equation}
        \eta_t=\frac{\delta H}{\delta \zeta},\quad
        \zeta_t=-\frac{\delta H}{\delta \eta}. 
\end{equation}

In the case of a free surface the results can be extended as in \cite{bib5,bib6,bib9, bib10}.

\section{THE LONG-WAVE APPROXIMATION}

As the vortex sheet provides a physically unrealistic representation it can be eliminated simply be letting $\kappa=\kappa_1=\kappa_2$ giving the Hamiltonian:
\begin{multline}
H\Big(\eta(x),\xi(x)\Big)=\frac{1}{2}\int\limits_{\mathbb{R}} \xi G_1(\eta) B^{-1}G_2(\eta)\xi \,dx
- \frac{1}{2}\rho_1\rho_2(\gamma_1-\gamma_2)^2\int\limits_{\mathbb{R}}  \eta\eta_x   B^{-1}\eta\eta_x  \,dx
-\int\limits_{\mathbb{R}} (\gamma_1\eta+\kappa)\xi\eta_x \,dx\\+\rho_2(\gamma_1-\gamma_2)\int\limits_{\mathbb{R}} \eta\eta_x B^{-1}G_1(\eta)\xi\,dx
+\frac{1}{6}(\rho_1\gamma_1^2 -\rho_2\gamma_2^2)\int\limits_{\mathbb{R}} \eta^3dx+\frac{1}{2}\Big(g(\rho_1-\rho_2 )+(\rho_1\gamma_1-\rho_2\gamma_2)\kappa\Big)\int\limits_{\mathbb{R}} \eta^2 dx.
\end{multline}

The Dirichlet-Neumann operators can be expanded in terms of orders of $\eta$ as
\begin{equation}
G_i(\eta)=\sum_{j=0}^{\infty} G_{ij}(\eta).
\end{equation}

As $\eta,\xi$ are periodic it can be shown that
\begin{equation}   
    G_1(\eta)\!=\!D\tanh(h_1 D)+D \eta D -D\tanh(h_1 D)  \eta D\tanh(h_1D)\!+\!\mathcal{O}(\eta^2)
\end{equation}
and
\begin{equation}   
    G_2(\eta)\!=\!D\tanh(h_2 D)- D \eta D +D\tanh(h_2 D) \eta D\tanh(h_2 D)\!+\!\mathcal{O}(\eta^2)
\end{equation}
where $D$ is a Fourier multiplier equivalent to the operation $-i\partial/\partial x$. $B(G(\eta))$ can hence also be expanded.

The following scaling is performed:
\begin{equation}   
    (x,y)\rightarrow h_2(x,y),\quad
    ({\bf{u}}_1,{\bf{u}}_2)\rightarrow\sqrt{ g h_2}({\bf{u}}_1,{\bf{u}}_2),\quad
    \eta\rightarrow a\eta
\end{equation}
where $({\bf{u}}_1,{\bf{u}}_2)$ are the velocity vectors
and $a$ is the wave amplitude.

Small arbitrary constant parameters $\varepsilon,\delta\in{\mathbb{R}}$ are introduced:
\begin{equation}
\varepsilon=\frac{a}{h_2},\quad\delta=\frac{h_2}{\lambda}.
\end{equation}

From the kinematic boundary condition $\eta,\xi \sim\mathcal{O}(\varepsilon)$.
 $\lambda\rightarrow h_2\lambda$ and so $\lambda^{-1}\sim\mathcal{O}(\delta)$ giving the long-wave regime. As $D$ is equivalent to the wavenumber $k=2\pi/\lambda$ then $ D\sim \mathcal{O}(\delta)$.

So the expanded, nondimensionalised and scaled Dirichlet-Neumann operators and  $B^{-1}$ operator can be written as
\begin{align}   
    G_1(\eta)&=\delta^2 \Big( h_1 D^2+\varepsilon D \eta D\Big)-\delta^4\Big(\frac{1}{3} h_1^3 D^4+\varepsilon  h_1^2 D^2 \eta D^2\Big )+\delta^6\Big(\frac{2}{15}h_1^5 D^6\Big) 
+\mathcal{O}(\delta^8,\varepsilon\delta^6,\varepsilon^2\delta^4), \\
   G_2(\eta)&=\delta^2\Big( h_2 D^2-\varepsilon D \eta D\Big)+\delta^4\Big(-\frac{1}{3}h_2^3 D^4+\varepsilon  h_2^2 D^2 \eta D^2 \Big)+\delta^6\Big(\frac{2}{15}h_2^5 D^6 \Big)
+\mathcal{O}(\delta^8,\varepsilon\delta^6,\varepsilon^2\delta^4)
\end{align}
and
\begin{multline}
{B}^{-1}=\frac{1}{\delta^2 (\rho_2 h_1+\rho_1 h_2)}D^{-1}\bigg\{1-\varepsilon\frac{\rho_2 -\rho_1}{\rho_2 h_1+\rho_1 h_2} \eta+\varepsilon^2\frac{(\rho_2 -\rho_1 )^2}{(\rho_2 h_1+\rho_1 h_2)^2} \eta^2 \\
+\delta^2\bigg[\frac{1}{3}\frac{\rho_2  h_1^3+\rho_1 h_2^3}{\rho_2 h_1+\rho_1 h_2}D^2-\frac{1}{3}\varepsilon \frac{(\rho_2 -\rho_1 )(\rho_2  h_1^3+\rho_1 h_2^3)}{(\rho_2 h_1+\rho_1 h_2)^2} \eta D^2
 -\frac{1}{3} \varepsilon\frac{(\rho_2 -\rho_1 )(\rho_2  h_1^3+\rho_1 h_2^3)}{(\rho_2 h_1+\rho_1 h_2)^2}D^2 \eta +\varepsilon \frac{\rho_2 h_1^2 -\rho_1 h_2^2}{\rho_2 h_1+\rho_1 h_2}D \eta D\bigg]\\
-\delta^4\bigg[\frac{2}{15}\frac{\rho_2  h_1^5+\rho_1  h_2^5}{\rho_2 h_1+\rho_1 h_2}D^4  -\frac{1(\rho_2  h_1^3+\rho_1 h_2^3)^2}{9(\rho_2 h_1+\rho_1 h_2)^2}D^4\bigg]+\mathcal{O}(\delta^6,\varepsilon\delta^4,\varepsilon^2\delta^2,\varepsilon^3)\bigg\}D^{-1}
\end{multline}
respectively noting the introduction of new notation such as bars or asterisks has been avoided.

The approximation to $\mathcal{O}(\delta^{6},\varepsilon\delta^4,\varepsilon^2\delta^2,\varepsilon^3)$ can now be calculated as
\begin{multline}
H(\eta,\xi)=\frac{1}{2}\delta^4\bigg(\frac{h_1 h_2}{\rho_2 h_1+\rho_1 h_2}\bigg)\int\limits_{\mathbb{R}} \xi   
{D}^2\xi  dx
+\frac{1}{2}\varepsilon\delta^4\bigg(\frac{\rho_1 h_2^2-\rho_2 h_1^2}{(\rho_2 h_1+\rho_1 h_2)^2}\bigg)\int\limits_{\mathbb{R}} \xi   
D \eta  D\xi  dx\\
-\frac{1}{2}\delta^6 \bigg(\frac{1}{3}\frac{h_1^2 h_2^2(\rho_2   h_2+\rho_1 h_1)}{(\rho_2 h_1+\rho_1 h_2)^2}\bigg)\int\limits_{\mathbb{R}} \xi   
 {D}^4\xi  dx
-\varepsilon\delta^2\kappa\int\limits_{\mathbb{R}} \xi \eta_x dx
+\frac{1}{2}\varepsilon^2\delta^2 \bigg(\frac{\gamma_1\rho_1 h_2+\rho_2\gamma_2 h_1}{ \rho_2 h_1+\rho_1 h_2}\bigg)\int\limits_{\mathbb{R}} \eta^2 \xi_xdx\\
+\frac{1}{6}\varepsilon^3(\rho_1\gamma_1^2 -\rho_2\gamma_2^2)\int\limits_{\mathbb{R}} \eta^3 dx
+\frac{1}{2}\varepsilon^2\Big(g(\rho_1-\rho_2 )+(\rho_1\gamma_1-\rho_2\gamma_2)\kappa\Big)\int\limits_{\mathbb{R}}\eta^2 dx.
\end{multline}

\section{THE KDV APPROXIMATION}
Now the assumption is made that
$\mathcal{O}(\varepsilon)\sim\mathcal{O}(\delta^2)$, with a view to establishing balance between nonlinearity and dispersion when obtaining solitary wave solutions, and we can write
\begin{multline}
H(\eta,\xi)=\frac{1}{2}\delta^4\alpha_1\int\limits_{\mathbb{R}} \xi   
{D}^2\xi  dx
+\frac{1}{2}\delta^6\alpha_3\int\limits_{\mathbb{R}} \eta\big(\xi_x\big)^2 dx
-\frac{1}{2}\delta^6 \alpha_2\int\limits_{\mathbb{R}} \xi   
 {D}^4\xi  dx
-\delta^4\kappa\int\limits_{\mathbb{R}} \xi \eta_x dx\\
+\frac{1}{2}\delta^6 \alpha_4\int\limits_{\mathbb{R}} \eta^2 \xi_xdx
+\frac{1}{6}\delta^6\alpha_6\int\limits_{\mathbb{R}} \eta^3 dx
+\frac{1}{2}\delta^4\alpha_5\int\limits_{\mathbb{R}} \eta^2 dx  
\end{multline}
where the following constants have been introduced:
\begin{align}
\alpha_1&=\dfrac{h_1 h_2}{\rho_2 h_1+\rho_1 h_2},          &  \alpha_ 2&=\frac{1}{3}\dfrac{h_1^2 h_2^2(\rho_2   h_2+\rho_1 h_1)}{(\rho_2 h_1+\rho_1 h_2)^2},              &  \alpha_3&=\dfrac{\rho_1 h_2^2-\rho_2 h_1^2}{(\rho_2 h_1+\rho_1 h_2)^2},\\
\alpha_4&=\dfrac{\gamma_1\rho_1 h_2+\rho_2\gamma_2 h_1}{ \rho_2 h_1+\rho_1 h_2},         &  \alpha_5&=g(\rho_1-\rho_2 )+(\rho_1\gamma_1-\rho_2\gamma_2)\kappa,  &  \alpha_6&=\rho_1\gamma_1^2 -\rho_2\gamma_2^2.
\end{align}

This gives the following (non-canonical) equations of motion,
\begin{equation}   
        \frac{\partial \eta}{\partial t}
=
-\alpha_1 \xi_{xx} 
-\delta^2\alpha_3 \big(\eta\xi_x\big)_x
-\delta^2\alpha_2\xi_{xxxx}
-\kappa\eta_x
-\delta^2 \alpha_4 \eta\eta_x
\end{equation}
and
\begin{equation}   
\label{EOM_2}
        \frac{\partial \xi}{\partial t}=-\frac{\delta^2}{2}\alpha_3\big(\xi_x\big)^2
-\kappa\xi_x
-\delta^2 \alpha_4 \eta\xi_x
-\frac{1}{2}\delta^2\alpha_6 \eta^2 -\alpha_5 \eta
+\Gamma\chi. 
\end{equation}

Next, we get the partial derivative of (\ref{EOM_2}) with respect to $x$ and it is noted that a term $g-2\omega\kappa$ appears which can be simplified by taking
\begin{equation}
g\approx 9.8 ms^{-2}, \quad\omega\approx7.3\times10^{-5}s^{-1},\quad\kappa\approx 1m s^{-1} 
\end{equation}
and thus clearly $g>>2\omega\kappa$.

We introduce
\begin{equation}
u=\xi_x 
\end{equation}
and use a Galilean shift
 \begin{equation}
(x,t)\rightarrow(x-\kappa t,t)=(X,T)
\end{equation}
giving
\begin{equation} 
        \eta_T+\alpha_1 u_X  +\delta^2\alpha_2 u_{XXX}
+\delta^2\alpha_3 (u\eta)_X 
+\delta^2 \alpha_4 \eta\eta_X=0
\end{equation}
and
\begin{equation} 
u_T-\Gamma\alpha_1 u_X
+g(\rho_1-\rho_2 ) \eta_X
-\delta^2\Gamma\alpha_2 u_{XXX}+\delta^2\alpha_3uu_X
+\delta^2 \Big(\alpha_4 (u\eta)_X 
-\Gamma\alpha_3 (u\eta)_X 
+\alpha_6 \eta \eta_X
-\Gamma \alpha_4 \eta\eta_X\Big)=0. 
\end{equation}

The linearised equations are hence
\begin{equation}   
        \eta_T+\alpha_1 u_X  =0, \quad
        u_T-\Gamma\alpha_1 u_X
+g(\rho_1-\rho_2 ) \eta_X=0. 
\end{equation}

Periodicity of $\eta$ and $u$ means we can write
\begin{equation}   
        \eta(X,T)=\eta_0e^{i(kX-\Omega(k) T)},\quad
        u(X,T)=u_0e^{i(kX-\Omega(k)T)} 
\end{equation}
where $k$ is the wavenumber, $\Omega(k)$ is the angular frequency and both are related via
\begin{equation}
c(k)=\frac{\Omega(k)}{k}
\end{equation}
where the wavespeed $c(k)$ is that observed by an observer moving at a speed $\kappa$ along $y=0$.

We hence have equations
\begin{equation}  
        -ick\eta+ i\alpha_1ku 
=0
\end{equation}
and
\begin{equation}  
        -icku
 +ig(\rho_1-\rho_2 ) k\eta
- i\Gamma\alpha_1ku
=0 .
\end{equation}

This has solutions  for observers moving with the flow as
\begin{equation}
c=\frac{1}{2}\bigg(-\alpha_1\Gamma\pm\sqrt{\alpha_1^2\Gamma^2+4\alpha_1 g(\rho_1-\rho_2)}\bigg) 
\end{equation}
and for stationary observers as
\begin{equation}
c+\kappa=\frac{1}{2}\bigg(-\alpha_1\Gamma\pm\sqrt{\alpha_1^2\Gamma^2+4\alpha_1 g(\rho_1-\rho_2)}\bigg). 
\end{equation}

Starting with the leading approximation
\begin{equation}
        u=\frac{c}{\alpha_1}\eta 
\end{equation}
the aim is to find a KdV type equation of the form
\begin{equation}
        \eta_T+c\eta_X+\delta^2\Big(C_1\eta\eta_X+C_2\eta_{XXX}\Big)=0 
\end{equation}
for some constants $C_1, C_2$, that is an equation that has a nonlinear and a dispersive component.

By considering a Johnson-type transformation \cite{bib11}
\begin{equation}
        u=\frac{c}{\alpha_1}\eta+\delta^2\Big(\mu\eta^2+\sigma\eta_{XX}\Big) 
\end{equation}
we calculate
\begin{equation}   
        \sigma=-\dfrac{c\alpha_2\bigg(1-\frac{\alpha_1 \Gamma}{c}\bigg)}{\alpha_1^2\bigg(2+\frac{\alpha_1 \Gamma}{c} \bigg)}
\end{equation}
and 
\begin{equation}   
\mu =\dfrac{\alpha_1\alpha_4 (c-\alpha_1\Gamma)
-\alpha_3c (c+2\alpha_1\Gamma)    +\alpha_1^2\alpha_6  }{2\alpha_1^2(2c+\Gamma\alpha_1)} 
\end{equation}
giving the KdV approximation \cite{bib4}
\begin{equation}   \label{kdv_p}
        \eta_T+ c\eta_X
+\delta^2\Bigg(   \frac{\alpha_1^2\alpha_6+3\alpha_3 c^2+3\alpha_1\alpha_4 c}{\alpha_1(2c+\Gamma\alpha_1)}   \Bigg)\eta \eta_X 
+\delta^2\Bigg(\frac{c^2\alpha_2 }{\alpha_1(2c+\alpha_1\Gamma)}\Bigg)\eta_{XXX}
=0 .
\end{equation}

The KdV regime is one of the most important propagation regimes for water waves, where stable nonlinear solitary waves (solitons) are formed, see also \cite{bib12, bib13, bib14}.  However, there are various propagation regimes and many other situations are possible, including $\delta^2 \ll \varepsilon.$ This is when the wavelength is very large in comparison to $h_i$.    In such case the contribution from the $\delta^2$ terms is not significant and can be neglected. Instead of a KdV equation the relevant model is the dispersionless Burgers equation ($\partial_{\tau}=\partial_T+c \partial_X$)

\begin{equation} \label{B} 
\eta_{\tau}+\varepsilon \Bigg(   \frac{\alpha_1^2\alpha_6+3\alpha_3 c^2+3\alpha_1\alpha_4 c}{\alpha_1(2c+\Gamma\alpha_1)}   \Bigg)\eta \eta_X=0.  
\end{equation}

Such an equation does not support globally smooth solutions, i.e. the solutions always form a vertical slope and break. Such wave-breaking phenomenon is well known for internal waves in the ocean. This is a mechanism that causes mixing in the deep ocean, with implications for biological productivity and sediment transport.

\section{SOLITARY WAVE SOLUTION}

We seek to obtain a KdV equation in the \emph{standard} format:
\begin{equation}
        \eta_T
+6\eta \eta_X 
+\eta_{XXX}
=0
\end{equation}
for which a solution, $\eta(X,T)=2\kappa^2 \text{sech}^2\big(\kappa (X-4\kappa^2 T)\big)$, is already known. 

We transform $T$ using
\begin{equation}
X\rightarrow X+cT 
\end{equation}
and scale as follows \cite{bib12}:
\begin{equation}  
    \eta\rightarrow \alpha \eta, \quad
    X\rightarrow\beta X,\quad
    T\rightarrow \gamma T.
\end{equation}

Introducing  
\begin{equation}  
A=    \frac{\alpha_1^2\alpha_6+3\alpha_3 c^2+3\alpha_1\alpha_4 c}{\alpha_1(2c+\Gamma\alpha_1)},    \quad B= \frac{c^2\alpha_2 }{\alpha_1(2c+\alpha_1\Gamma)}
\end{equation}
the KdV is hence written
\begin{equation}  
        \eta_T
+\delta^2A\eta \eta_X 
+\delta^2B\eta_{XXX}
=0
\end{equation}
and the relation
\begin{equation}
\frac{\gamma}{\beta}=\frac{6}{\delta^2\alpha A}=\frac{\beta^2}{\delta^2B}  
\end{equation}
is noted. 

Choosing 
\begin{equation}
\gamma=\frac{1}{\delta^2}
\end{equation}
we can establish that 
\begin{equation}
\alpha= \frac{6B^{\frac{1}{3}}}{A},\quad\beta=B^{\frac{1}{3}}
\end{equation}
giving the solitary wave solution of \eqref{kdv_p}
\begin{equation}   
        \eta(X,T)=\frac{12\kappa ^2 B^{1/3}}{A} \text{sech}^2 \bigg\{\frac{\kappa}{B^{1/3}}\Big(X+(c-4\delta^2\kappa^2B^{1/3}) T\Big)\bigg\}
\end{equation}
which describes a solitary crest of amplitude $12\kappa ^2 B^{1/3}/A$ moving with speed $c-4\delta^2\kappa^2B^{1/3}$. The correction to the speed $c$ is $-4\delta^2\kappa^2B^{1/3},$ which is related to the amplitude through $\kappa$.

% Acknowledgement
\section{ACKNOWLEDGMENTS}
The author is grateful to Prof. A. Constantin, Dr R. Ivanov and Dr C.I. Martin for many valuable discussions and to Prof. M. Todorov for organising the Ninth Conference of the Euro-American Consortium for Promoting the Application of Mathematics in Technical and Natural Sciences. Financial support from the Erwin Schr\"odinger International Institute for Mathematics and Physics (ESI), Vienna (Austria) for participation in the Research in Teams Project {\it Hamiltonian approach to modelling geophysical waves and currents with impact on natural hazards} is acknowledged. The author is supported by the Fiosraigh scholarship program at Dublin Institute of Technology (Ireland). 

% References

\nocite{*}
\bibliographystyle{aipnum-cp}%
\bibliography{aip_amitans17_revision_acompelli}%

%Control: year  (1) truncated
\begin{thebibliography}{14}%
\makeatletter
\providecommand \@ifxundefined [1]{%
 \@ifx{#1\undefined}
}%
\providecommand \@ifnum [1]{%
 \ifnum #1\expandafter \@firstoftwo
 \else \expandafter \@secondoftwo
 \fi
}%
\providecommand \@ifx [1]{%
 \ifx #1\expandafter \@firstoftwo
 \else \expandafter \@secondoftwo
 \fi
}%
\providecommand \natexlab [1]{#1}%
\providecommand \enquote  [1]{``#1''}%
\providecommand \bibnamefont  [1]{#1}%
\providecommand \bibfnamefont [1]{#1}%
\providecommand \citenamefont [1]{#1}%
\providecommand \href@noop [0]{\@secondoftwo}%
\providecommand \href [0]{\begingroup \@sanitize@url \@href}%
\providecommand \@href[1]{\@@startlink{#1}\@@href}%
\providecommand \@@href[1]{\endgroup#1\@@endlink}%
\providecommand \@sanitize@url [0]{\catcode `\$12\catcode `\&12\catcode
  `\#12\catcode `\^12\catcode `\_12\catcode `\%12\relax}%
\providecommand \@@startlink[1]{}%
\providecommand \@@endlink[0]{}%
\providecommand \url  [0]{\begingroup\@sanitize@url \@url }%
\providecommand \@url [1]{\endgroup\@href {#1}{\urlprefix }}%
\providecommand \urlprefix  [0]{URL }%
\providecommand \Eprint [0]{\href }%
\providecommand \doibase [0]{http://dx.doi.org/}%
\providecommand \selectlanguage [0]{\@gobble}%
\providecommand \bibinfo  [0]{\@secondoftwo}%
\providecommand \bibfield  [0]{\@secondoftwo}%
\providecommand \translation [1]{[#1]}%
\providecommand \BibitemOpen [0]{}%
\providecommand \bibitemStop [0]{}%
\providecommand \bibitemNoStop [0]{.\EOS\space}%
\providecommand \EOS [0]{\spacefactor3000\relax}%
\providecommand \BibitemShut  [1]{\csname bibitem#1\endcsname}%
\let\auto@bib@innerbib\@empty
%</preamble>
\bibitem [{\citenamefont {Alford}\ \emph {et~al.}(2015)\citenamefont {Alford}
  \emph {et~al.}}]{bib1}%
  \BibitemOpen
  \bibfield  {author} {\bibinfo {author} {\bibfnamefont {M.}~\bibnamefont
  {Alford}} \emph {et~al.},\ }\href {\doibase doi:10.1038/nature14399}
  {\bibfield  {journal} {\bibinfo  {journal} {Nature}\ }\textbf {\bibinfo
  {volume} {521}},\ \unskip\ \bibinfo {pages} {65--69} (\bibinfo {year}
  {2015})}\BibitemShut {NoStop}%
\bibitem [{\citenamefont {Craig}, \citenamefont {Guyenne},\ and\ \citenamefont
  {Kalisch}(2005)}]{bib2}%
  \BibitemOpen
  \bibfield  {author} {\bibinfo {author} {\bibfnamefont {W.}~\bibnamefont
  {Craig}}, \bibinfo {author} {\bibfnamefont {P.}~\bibnamefont {Guyenne}}, \
  and\ \bibinfo {author} {\bibfnamefont {H.}~\bibnamefont {Kalisch}},\
  }\href@noop {} {\bibfield  {journal} {\bibinfo  {journal} {Comm. Pure Appl.
  Math.}\ }\textbf {\bibinfo {volume} {58(12)}},\ \unskip\ \bibinfo {pages}
  {1587--1641} (\bibinfo {year} {2005})}\BibitemShut {NoStop}%
\bibitem [{\citenamefont {Compelli}(2017)}]{bib3}%
  \BibitemOpen
  \bibfield  {author} {\bibinfo {author} {\bibfnamefont {A.}~\bibnamefont
  {Compelli}},\ }\href@noop {} {\bibfield  {journal} {\bibinfo  {journal} {Wave
  Motion}\ }\textbf {\bibinfo {volume} {54}},\ \unskip\ \bibinfo {pages}
  {115--124} (\bibinfo {year} {2017})}\BibitemShut {NoStop}%
\bibitem [{\citenamefont {Compelli}\ and\ \citenamefont {Ivanov}(2017)}]{bib4}%
  \BibitemOpen
  \bibfield  {author} {\bibinfo {author} {\bibfnamefont {A.}~\bibnamefont
  {Compelli}}\ and\ \bibinfo {author} {\bibfnamefont {R.}~\bibnamefont
  {Ivanov}},\ }\href@noop {} {\bibfield  {journal} {\bibinfo  {journal} {J.
  Math. Fluid Mech.}\ }\textbf {\bibinfo {volume} {19}},\ \unskip\ \bibinfo
  {pages} {329--344} (\bibinfo {year} {2017})},\ \bibinfo {note}
  {arXiv:1611.06581 [physics.flu-dyn]}\BibitemShut {NoStop}%
\bibitem [{\citenamefont {Ivanov}(2017)}]{bib5}%
  \BibitemOpen
  \bibfield  {author} {\bibinfo {author} {\bibfnamefont {R.}~\bibnamefont
  {Ivanov}},\ }\href@noop {} {\bibfield  {journal} {\bibinfo  {journal}
  {Nonlinear Analysis: Real World Applications}\ }\textbf {\bibinfo {volume}
  {34}},\ \unskip\ \bibinfo {pages} {316--334} (\bibinfo {year} {2017})},\
  \bibinfo {note} {corrigendum vol. 36 (2017) 115, arXiv:1702.01441
  [physics.flu-dyn]}\BibitemShut {NoStop}%
\bibitem [{\citenamefont {Constantin}\ and\ \citenamefont
  {Johnson}(2015)}]{bib6}%
  \BibitemOpen
  \bibfield  {author} {\bibinfo {author} {\bibfnamefont {A.}~\bibnamefont
  {Constantin}}\ and\ \bibinfo {author} {\bibfnamefont {R.}~\bibnamefont
  {Johnson}},\ }\href@noop {} {\bibfield  {journal} {\bibinfo  {journal}
  {Geophysical and Astrophysical Fluid Dynamics}\ }\textbf {\bibinfo {volume}
  {109}},\ \unskip\ \bibinfo {pages} {311--358} (\bibinfo {year}
  {2015})}\BibitemShut {NoStop}%
\bibitem [{\citenamefont {Constantin}, \citenamefont {Ivanov},\ and\
  \citenamefont {Prodanov}(2008)}]{bib7}%
  \BibitemOpen
  \bibfield  {author} {\bibinfo {author} {\bibfnamefont {A.}~\bibnamefont
  {Constantin}}, \bibinfo {author} {\bibfnamefont {R.}~\bibnamefont {Ivanov}},
  \ and\ \bibinfo {author} {\bibfnamefont {E.}~\bibnamefont {Prodanov}},\
  }\href@noop {} {\bibfield  {journal} {\bibinfo  {journal} {J. Math. Fluid
  Mech.}\ }\textbf {\bibinfo {volume} {10}},\ \unskip\ \bibinfo {pages}
  {224--237} (\bibinfo {year} {2008})}\BibitemShut {NoStop}%
\bibitem [{\citenamefont {Wahlen}(2007)}]{bib8}%
  \BibitemOpen
  \bibfield  {author} {\bibinfo {author} {\bibfnamefont {E.}~\bibnamefont
  {Wahlen}},\ }\href@noop {} {\bibfield  {journal} {\bibinfo  {journal} {Lett.
  Math. Phys.}\ }\textbf {\bibinfo {volume} {79}},\ \unskip\ \bibinfo {pages}
  {303--315} (\bibinfo {year} {2007})}\BibitemShut {NoStop}%
\bibitem [{\citenamefont {Constantin}\ and\ \citenamefont
  {Ivanov}(2015)}]{bib9}%
  \BibitemOpen
  \bibfield  {author} {\bibinfo {author} {\bibfnamefont {A.}~\bibnamefont
  {Constantin}}\ and\ \bibinfo {author} {\bibfnamefont {R.}~\bibnamefont
  {Ivanov}},\ }\href@noop {} {\bibfield  {journal} {\bibinfo  {journal} {Phys.
  Fluids}\ }\textbf {\bibinfo {volume} {27}} (\bibinfo {year} {2015})},\
  \bibinfo {note} {086603}\BibitemShut {NoStop}%
\bibitem [{\citenamefont {Constantin}, \citenamefont {Ivanov},\ and\
  \citenamefont {Martin}(2016)}]{bib10}%
  \BibitemOpen
  \bibfield  {author} {\bibinfo {author} {\bibfnamefont {A.}~\bibnamefont
  {Constantin}}, \bibinfo {author} {\bibfnamefont {R.}~\bibnamefont {Ivanov}},
  \ and\ \bibinfo {author} {\bibfnamefont {C.~I.}\ \bibnamefont {Martin}},\
  }\href {\doibase 10.1007/s00205-016-0990-2} {\bibfield  {journal} {\bibinfo
  {journal} {Arch. Rational Mech. Anal.}\ }\textbf {\bibinfo {volume} {221}},\
  \unskip\ \bibinfo {pages} {1417--1447} (\bibinfo {year} {2016})},\ \bibinfo
  {note} {arXiv:math-ph/0610014}\BibitemShut {NoStop}%
\bibitem [{\citenamefont {Johnson}(2002)}]{bib11}%
  \BibitemOpen
  \bibfield  {author} {\bibinfo {author} {\bibfnamefont {R.}~\bibnamefont
  {Johnson}},\ }\href@noop {} {\bibfield  {journal} {\bibinfo  {journal} {J.
  Fluid Mech.}\ }\textbf {\bibinfo {volume} {455}},\ \unskip\ \bibinfo {pages}
  {63--82} (\bibinfo {year} {2002})}\BibitemShut {NoStop}%
\bibitem [{\citenamefont {Johnson}(1997)}]{bib12}%
  \BibitemOpen
  \bibfield  {author} {\bibinfo {author} {\bibfnamefont {R.}~\bibnamefont
  {Johnson}},\ }in\ \href@noop {} {\emph {\bibinfo {booktitle} {A Modern
  Introduction to the Mathematical Theory of Water Waves}}}\ (\bibinfo
  {publisher} {Cambridge University Press},\ \bibinfo {address} {Cambridge},\
  \bibinfo {year} {1997})\BibitemShut {NoStop}%
\bibitem [{\citenamefont {Ivanov}(2007)}]{bib13}%
  \BibitemOpen
  \bibfield  {author} {\bibinfo {author} {\bibfnamefont {R.}~\bibnamefont
  {Ivanov}},\ }\href@noop {} {\bibfield  {journal} {\bibinfo  {journal}
  {Philos. Trans. Roy. Soc.: Ser. A}\ }\textbf {\bibinfo {volume} {365}},\
  \unskip\ \bibinfo {pages} {2267--2280} (\bibinfo {year} {2007})},\ \bibinfo
  {note} {arXiv:0707.1839 [nlin.SI]}\BibitemShut {NoStop}%
\bibitem [{\citenamefont {Henry}\ and\ \citenamefont {Ivanov}(2014)}]{bib14}%
  \BibitemOpen
  \bibfield  {author} {\bibinfo {author} {\bibfnamefont {D.}~\bibnamefont
  {Henry}}\ and\ \bibinfo {author} {\bibfnamefont {R.}~\bibnamefont {Ivanov}},\
  }\href@noop {} {\bibfield  {journal} {\bibinfo  {journal} {Discrete Contin.
  Dyn. Syst.-A}\ }\textbf {\bibinfo {volume} {34}},\ \unskip\ \bibinfo {pages}
  {3025--3034} (\bibinfo {year} {2014})},\ \bibinfo {note} {arXiv:1402.0537
  [nlin.SI]}\BibitemShut {NoStop}%
\end{thebibliography}%

\end{document}